\documentclass[twocolumn,showpacs,preprintnumbers,amsmath,amssymb,superscriptaddress,prl]{revtex4}

\usepackage{color}
\usepackage{graphicx}
\usepackage{dcolumn}
\usepackage{bm}

\begin{document}
\newcommand{\Cs}{C$_{60}$}
\newcommand{\Ct}{C$_{120}$}
\newcommand{\Ca}{C$_{70}$}
\newcommand{\Cb}{C$_{82}$}
\newcommand{\Sc}{Sc@C$_{82}$}
\newcommand{\Cc}{C$_{86}$}
\newcommand{\Csm}{C$_{60}^-$}
\newcommand{\Csmm}{C$_{60}^{2-}$}
\newcommand{\Csmmn}{(C$_{60})_2^{2-}$}
\newcommand{\Csmmb}{(C$_{60})_2^{-}$}
\newcommand{\Csmmm}{C$_{60}^{3-}$}
\newcommand{\Csmmmmm}{C$_{60}^{5-}$}
\newcommand{\Csn}{C$_{60}^{n-}$}
\newcommand{\tu}{t$_{1u}$}
\newcommand{\tg}{t$_{2g}$}
\newcommand{\hu}{h$_{u}$}
\newcommand{\Ch}{C$_{2h}$}
\newcommand{\C}{C$_{2}$}
\newcommand{\Ci}{C$_{i}$}
\newcommand{\Ih}{I$_{h}$}
\newcommand{\Dh}{D$_{2h}$}
\newcommand{\kb}{k$_{b}$}
\newcommand{\m}{\mu_{B}}

\title{Modelling spin qubits in carbon peapods}

\author{Ling Ge}
\email{ling.ge@materials.ox.ac.uk}
\affiliation{Department of Materials, University of Oxford, Oxford OX1 3PH, United Kingdom}
\author{Barbara Montanari}
\affiliation{STFC Rutherford Appleton Laboratory, Didcot, Oxfordshire OX11 0QX, United Kingdom}
\author{John H. Jefferson}
\affiliation{Sensors and Electronics Division, QinetiQ, St Andrews Road, Malvern, 
Worcs. WR14 3PS, United Kingdom}
\author{David G. Pettifor}
\affiliation{Department of Materials, University of Oxford, Oxford OX1 3PH, United Kingdom}
\author{Nicholas M. Harrison}
\affiliation{STFC Rutherford Appleton Laboratory, Didcot, Oxfordshire OX11 0QX, United Kingdom}
\affiliation{Department of Chemistry, Imperial College, London SW7 2AZ, United Kingdom}
\author{G. Andrew D. Briggs}
\affiliation{Department of Materials, University of Oxford, Oxford OX1 3PH, United Kingdom}

\date{\today}
             
\begin{abstract}

We have calculated electron spin interactions in chains of Sc@C$_{82}$~endohedral fullerenes in isolation and inserted into a semiconducting or metallic single-walled carbon nanotube to form a peapod. Using hybrid density functional theory (DFT), we find that the spin resides mainly on the fullerene cage, whether or not the fullerenes are in a nanotube. The spin interactions decay exponentially with fullerene separation, and the system can be described by a simple antiferromagnetic Heisenberg spin chain. A generalised Hubbard-Anderson model gives an exchange parameter $J$ and a Coulomb parameter $U$ in good agreement with the DFT values. Within the accuracy of the calculations, neither semiconducting nor metallic nanotubes affect the interactions between the fullerene electron spins.

\end{abstract}

\maketitle

Spin qubits have potential for controlled interactions~\cite{Los98} for quantum computing. Carbon is a candidate host for spin qubits because in $^{12}$C materials the small spin-orbit coupling and absence of hyperfine coupling ensures long spin coherence times. Carbon peapods~\cite{Smi98}, that is, single-walled carbon nanotubes (SWNT) containing fullerenes, have been proposed as hosts for spin-qubits~\cite{Ard03}. The fabrication of nanoscale electronic devices such as field effect transistors with carbon peapods containing various endohedral fullerenes is well established~\cite{Kit07}. When spin active metallic atoms such as Sc are incarcerated in a carbon cage, the system develops hybridized orbitals resulting in an unpaired electron delocalized across the fullerene cage~\cite{Mor05}. Here we report on detailed numerical simulations which establish the nature of the spin-spin interactions both between endohedral fullerenes and between fullerenes and nanotubes. The dominant interaction is of the Heisenberg form, which is known to be suitable for quantum computing (QC) in one-dimensional chains~\cite{Ben04}. Moreover, these results support this system to be promising for experimentally-proven QC protocols which allow chains formed of identical units to be controlled ``globally''~\cite{Fit07}, i.e. without the need to target individual units, leading to a scalable molecular quantum computer. 

\begin{figure}[h]
\begin{center}
    \includegraphics[width=8cm]{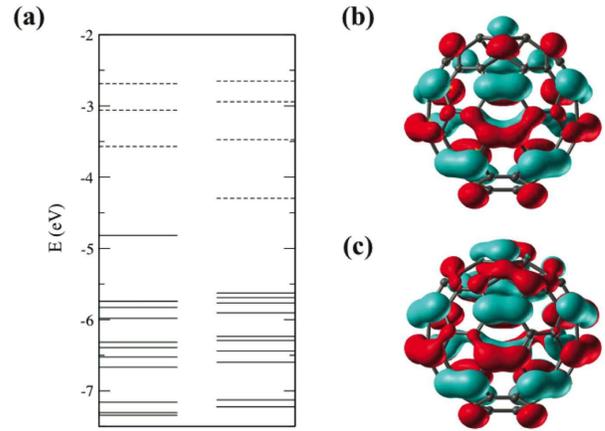}
  \caption{(Color online) (a) Eigenspectrum of relaxed \Sc~molecule. Left (right) hand side refers to spin-up (-down) electrons. Solid (dashed) lines refer to occupied (unoccupied) energy levels. (b) HOMO of relaxed \Sc. (c) (LUMO+1) of relaxed \Cb. Dark grey (red online) and light grey (blue online) lobes represent positive and negative phases.}
  \label{ScC82}     
  \end{center}
\end{figure}

Four fundamental problems need to be understood in order to demonstrate
well-defined qubits in carbon peapods as proposed in Ref.~\cite{Ben06}: (i) the
charge arrangement within the carbon peapods; (ii) the electron spin distribution;
(iii) the coupling between spin-qubits; (iv) the nature of the spin interactions between
fullerenes and nanotube. Density functional theory (DFT)
calculations have been reported on several model systems, such as SWNT
containing \Cs, K$_x$\Cs, Y@\Cs, \Cb~and La@\Cb~\cite{Cho03, Oka01,
  Oka03, Ota03, Dub04, Du03}. To date little work has been done on understanding the spin properties of peapods and the above mentioned issues (ii), (iii) and (iv) have not been addressed in detail thus far. 
We have computed the charge and spin distributions and the electronic structures within \Sc~chains and peapod structures. We find well-defined spin-1/2 qubits on the fullerenes, with strong evidence for a nearest-neighbor Heisenberg exchange interaction. We can tune this exchange interaction strength by controlling the inter-fullerene spacing during synthesis. In order to describe the influence on the spin-qubits localized on the fullerenes from propagating electrons or holes in the nanotube, we need to go beyond DFT to a model which is capable of describing the low-energy charge-spin excitations of the system. We conjecture a generic Hubbard-Anderson model, which has these properties, to estimate the low-energy spin interactions; in particular the Heisenberg exchange between spins along the fullerene chain and the Kondo exchange interaction between localized spins on the fullerenes and spins of propagating electrons or holes in the nanotube.

As model systems, we choose \Sc~in (14,7) (semiconducting) and (11,11) (metallic) SWNTs. Our calculations predict an exothermic encapsulation of \Sc~for both tubes. The repeat units containing one \Sc~molecule in the (14,7) and (11,11) peapods are 11.42~\AA~and 12.47~\AA, respectively~\cite{Sai98}. In order to study the exchange interaction, we use double unit cells containing two \Sc~molecules. The interwall separations between the (14,7) and (11,11) tubes and \Sc~are 3.35~\AA~(van der Waals distance) and 3.55~\AA, respectively. The relaxed structure of \Sc~is found to be in agreement with Ref.~\cite{Mor05} with a Sc-C distance of 2.26~\AA.  The
DFT calculations are performed with the hybrid exchange density
functional B3LYP~\cite{Bec88, Bec93, Lee88} as implemented in the CRYSTAL
package~\cite{CRYSTAL06}. The calculations reported here are
all-electron, i.e., with no shape approximation to the ionic potential or
electron charge density. The geometry optimizations are performed using the
algorithm proposed by Schlegel {\it et al}~\cite{Sch82}. The crystalline wave functions are
expanded in Gaussian basis sets of double valence quality (6-21G* for C and 864-11G* for Sc). Atomic charges are estimated using Mulliken population analysis~\cite{Mul55}. The system is modelled as a one dimensionally periodic array and reciprocal space sampling is performed on a Monkhorst-Pack grid containing 30 symmetry irreducible k-points which converges the total energy to within $10^{-4}$~eV per unit cell. 

Fig.\ref{ScC82} (a) shows the calculated electronic eigenspectrum for the
relaxed \Sc~molecule. Sc has three valence electrons and the ground state of
\Sc~ is found to be a spin-1/2 system. The unpaired electron occupies the highest occupied molecular orbital (HOMO) of \Sc, which constitutes the spin qubit.  The HOMO-LUMO gap is 0.530 eV. The separation between the HOMO and (HOMO-1) is 0.789 eV. Therefore, the HOMO is well separated from the energy levels above and below leading to a well-defined qubit. The \Sc~HOMO is delocalized across the fullerene cage as depicted in Fig.\ref{ScC82} (b), which is in agreement with the results found in Ref. \cite{Mor05, Lu00}. Furthermore, we establish that the HOMO of \Sc~is virtually identical to the (LUMO+1) of \Cb, as shown in Fig.\ref{ScC82} (c), whereas the lower lying orbitals are hybrids of Sc and \Cb. Thus Sc acts as a perfect donor to the \Cb~cage for the HOMO state. 

\begin{figure}[h]
\begin{center}
    \includegraphics[width=8cm]{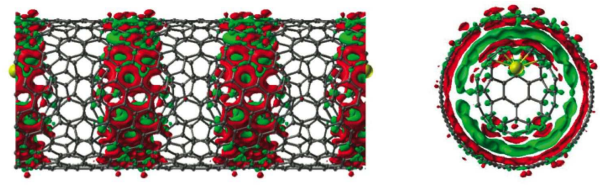}
  \caption{(Color online) Charge transfer in \Sc@(14,7) peapod. Values for dark grey (red online) and light grey (green online) surfaces are $\pm0.001~e$$/$\AA$^3$. Left (right) hand shows front (side) views. The atom colored in light grey (gold online) is Sc.}
  \label{ScpeapodCT}
  \end{center}
\end{figure}

Fig.~\ref{ScpeapodCT} shows the electronic charge rearrangement following the \Sc~encapsulation in the (14,7) nanotube. The charge depletion from the nanotube is concentrated around the fullerene sites. Similar qualitative results are obtained for the (11,11) nanotube, resembling that in La@C$_{82}$@(17,0)~\cite{Cho03}. Table 1 shows the charge and spin populations in \Sc, \Sc@(14,7) and
\Sc@(11,11) peapods. In \Sc, 1.64 electrons transfer from the Sc atom to the
\Cb~cage indicating a partially covalent Sc-cage bond.  In both \Sc@(14,7) and
\Sc@(11,11) peapods, electron transfer occurs from the nanotube and the Sc
atom to the \Cb~cage due to hybridization between the occupied states of
the nanotube and fullerenes. The charge transfer from the Sc atom to the
\Cb~cage in \Sc@(14,7) and \Sc@(11,11) is very similar to that of \Sc. The electron spin distribution is also similar as 97\% of the density resides on the \Cb~cage and only 3\% on the Sc atom. The shape of the spin density distribution in the peapods closely resembles that of the HOMO of \Sc~illustrated in Fig. \ref{ScC82} (b). The charge transfer and spin distribution are insensitive to encapsulation. 

\begin{center}
\medskip
{Table 1. Charge and spin populations in \Sc, \Sc@(14,7) and \Sc@(11,11) peapods}
\begin{tabular}{c c c c} \hline \hline
System & Component &   $q (e)$   &   $m (\m)$  \\ \hline
\Sc & Sc & 1.64 & 0.03 \\
& \Cb & -1.64 & 0.97 \\ \hline 
\Sc@(14,7) & Sc & 1.63 & 0.03 \\
& \Cb & -1.76 & 0.97  \\ 
& tube & 0.13 & 0.00 \\ \hline 
\Sc@(11,11) & Sc & 1.63 & 0.03 \\
& \Cb & -1.70 & 0.97  \\ 
& tube & 0.07 & 0.00 \\ \hline \hline
\end{tabular}
\end{center}

\begin{figure}[h]
\begin{center}
    \includegraphics[width=8cm]{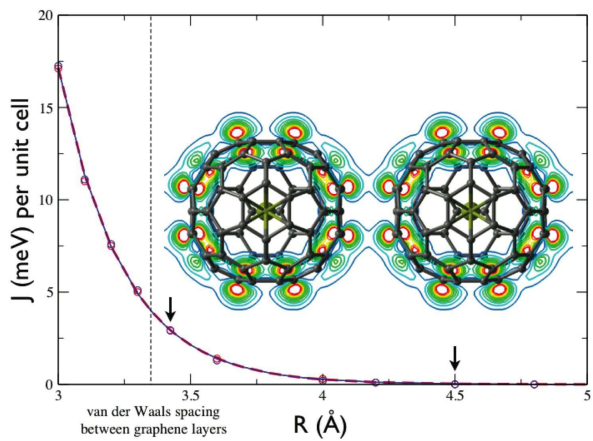}
  \caption{(Color online) Exchange interaction strength, $J$, as a function of inter-fullerene separation for a chain of \Sc~fullerenes. Empty circles refer to calculated results and are fitted using exponential decay law (lines) of the form $J_0e^{-\lambda (R-R_0)}$, where $J_0$ = 4.0~meV, $\lambda = 4.16$ \AA$^{-1}$, $R_0 = 3.35$~\AA. Dashed (red online) line refers to $J$ obtained from the DFT energy difference between FM and AF states. Solid (blue online) line refers to $J_{\rm eff}$ obtained from the Heisenberg model $4t^2/U_{\rm eff}$, where $U_{\rm eff}$ is fitted to be 0.412~eV from the value of $J$ at $R_0$. Arrows show discrete values of $R$ at which $J$ is calculated for peapods. Inset shows electron density contributed by highest occupied states of \Sc~chain.  Range of isovalues is 0-0.002~$e/$\AA$^3$.}
  \label{J}
  \end{center}
\end{figure}

In the predicted ground state configuration of the peapods, the spin direction alternates along the \Sc~chain; the corresponding configuration with parallel spins being higher in energy. We denote these configurations as antiferromagnetic (AF) and ferromagnetic (FM) states, respectively. Both states are found to be Mott insulators and have a total energy lower than the nonmagnetic state. For the \Sc@(14,7) peapod where the inter-fullerene spacing is 3.42~\AA, the energy difference between the nonmagnetic and FM states is 0.108~eV/cell. The exchange parameter $J$, defined as the
energy difference between FM and AF configurations, is 3~meV per cell (containing two spins). The behavior of $J$ as a function of the inter-fullerene separation, $R$, in a \Sc~chain is plotted in Fig.\ref{J}. Values of $J$ calculated for the peapods at discrete values of $R$, as indicated in Fig. \ref{J}, coincide with those obtained for the \Sc~chain within the accuracy of the present calculations. At these separations the inter-molecular spin interaction is therefore via direct exchange between fullerenes, with a negligible contribution from interactions via the nanotube. This inter-molecular coupling is much larger than the classical dipole coupling of N@C$_{60}$~\cite{Har02} and even larger than that computed for defective fullerenes with inter-cage links~\cite{Cha04}, ensuring $>10^3$ two-qubit gate operations within the decoherence time. This surprising result follows from the HOMOs in the \Sc~chain being very extended as illustrated in Fig.~\ref{J}. The $p_z$ orbitals on the closest C atoms belonging to adjacent molecules overlap in a $\sigma$-type fashion. This implies that the exchange interaction could be tuned by varying the separation between the fullerenes in peapods, for example by using functionalized fullerenes~\cite{Cha07}. Such high $J$ values are consistent with recent magnetic susceptibility experiments on \Sc~solids~\cite{Ito07}, which show AF Curie-Weiss temperature $\simeq$~300~K, consistent with $J = 17$~meV at $R = 3.08$~\AA~in Fig. 3, which corresponds to the measured lattice spacing.
\begin{figure}[h]
\begin{center}
    \includegraphics[width=8cm]{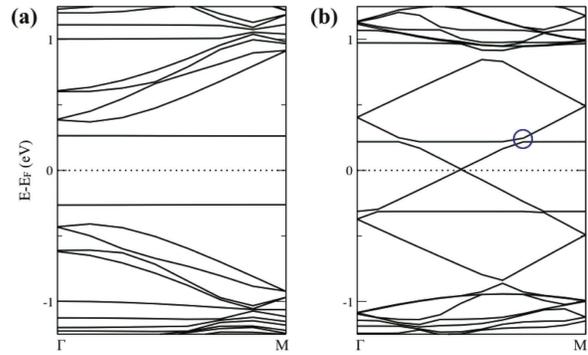}
  \caption{Spin-polarized band structures of the (a) \Sc@(14,7) and (b) \Sc@(11,11) peapods for the AF configuration. $E_F$ is the Fermi energy.}
  \label{Scpeapodband}
  \end{center}
\end{figure}

A generic Hubbard-Anderson model, going beyond DFT and capable of describing the low-energy spin interactions, is conjectured and the energy parameters of this model are estimated by direct comparison of its mean-field solutions with the DFT results. The Hubbard-Anderson model may be used directly to estimate the Heisenberg exchange between spins along the fullerene chain and the Kondo exchange interaction between localized spins on the fullerenes and spins of propagating electrons or holes in the nanotube. The Hamiltonian is
\begin{eqnarray}
H_{HA}  & = &  \sum_{lk\sigma}{\epsilon_{lk}}c_{lk\sigma}^\dagger c_{lk\sigma} + \sum_{k\sigma} {E_k a_{k\sigma}^\dagger a_{k\sigma}}   \\ \nonumber
&+& \sum_{lk\sigma}\gamma_{lk}[a_{k\sigma}^\dagger c_{lk\sigma} + h.c.] + U \sum_i n_{i\uparrow} n_{i\downarrow}
\end{eqnarray}
where $c_{lk\sigma}^\dagger$ is a creation operator for an electron in the nanotube with band $l$ with quasi-momentum $k$, $a_{k\sigma}^\dagger$ is a creation operator for an electron in the highest occupied fullerene band, $\gamma_{lk}$ is the hybridization parameter between fullerenes and nanotube, $U$ is the intra-fullerene Coulomb repulsion energy and $n_i$ is the number operator for the site $i$ of the \Sc~chain. 

The AF band structures of the peapods are plotted in Fig. 4. They display a very narrow but almost perfect cosine form for the highest occupied band~\cite{Sai98}. This is well described by a simple tight-binding model with the dispersion energy $\tilde{E}_k = E_0 - 2\tilde{t}{\cos}k$. It may be shown from an AF mean-field solution of the Hubbard model, that this dispersion corresponds to $E_k = E_0 - 2t{\cos}k$ in Eq. (1), where $t=\sqrt{\tilde{t}U}$, and $U$~$\simeq$~0.53~eV is obtained from the energy gap between centres of the highest occupied and lowest unoccupied bands in Fig. 4. This solution is also equivalent to a Stoner model for exchange $I = U$, where $J = 4t^2/U$~\cite{Pet95}. The behavior of the computed hopping parameter, $t$, as a function of $R$, in a \Sc~chain is fitted by an exponential decay law of the form $t_0e^{-\lambda (R-R_0)}$, where $t_0$ = 20.3~meV, $\lambda = 2.08$~\AA$^{-1}$, $R_0 = 3.35$~\AA, showing $t^2$ and $J$ scaling in the same way. For the values of $R$ in the (14,7) and (11,11) peapods, $t$ = 17~meV and 2~meV, respectively.  

An effective $U$ is fitted by fixing $J=4t^2/U_{\rm eff}$ within the Heisenberg model at $R_0$, giving $U_{\rm eff}$ =~0.412~eV, in good agreement with the DFT calculations. Fig. 3 shows that $J_{\rm eff}$ and $J$ are indistinguishable, implying approximately constant $U$ over the range of $R$ considered. $U$ deduced from the DFT band gap is consistent with the mean-field solution of the Hubbard-Anderson model and the total energy difference between FM and AF solutions. With $t\ll U$ for the peapods considered, we thus expect strongly correlated electron effects with well-defined spin-qubits along the \Sc~chain of the peapods and weak charge fluctuations. Therefore, the system is well characterized by an antiferromagnetic Heisenberg model.    

In the metallic peapod, we may also estimate the hybridization coupling parameters $\gamma_{lk}$ in Eq. (1) from the anticrossing gaps associated with the nanotube bands and a narrow fullerene band. For example, the anticrossing shown in the inset in Fig. \ref{Scpeapodband} (b) yields a coupling energy $\gamma_{lk}$=$\Delta E$/2=5 meV, where $\Delta E$ is the energy gap.

The weak interaction between electrons on the fullerenes and conduction electrons (or holes) in the nanotube, characterized by $\gamma_{lk}$ discussed above, will give rise to Kondo-like coupling between spins on the fullerenes and spins in the metallic nanotube. The energy scale for these couplings is given by $J_K \sim \gamma_{lk}^2/E_t$, where $E_t$ is a charge transfer energy gap, i.e. $E_t=E_0+U-E_F$ or $E_F-E_0$~\cite{Hew93}. This gives a typical $J_K\sim0.1$~meV for the (11,11) peapod.  

DFT calculations cannot resolve the difference in $J$ obtained for a \Sc~chain and the corresponding peapod structure even at large inter-fullerene separation. This is consistent with direct exchange dominating RKKY interactions. Charge fluctuations could be increased by either enhancing the hybridization interaction through a decrease in the fullerene-nanotube
separation, or by tuning the Fermi energy in metallic nanotubes to approach
the mixed valence regime in which the charge transfer energy tends to
zero. This would enhance both the Kondo coupling and the RKKY interaction, in
competition with the direct Heisenberg exchange.

In conlusion, the perfect doping mechanism of \Sc~leads to well-defined spin-qubits on the \Cb~cage in the peapods considered, coupled via antiferromagnetic Heisenberg exchange interaction. For the semiconducting case, the upper and lower Hubbard bands of the fullerene chain are little affected by the nanotube and occur entirely within the band gap of the nanotube, allowing excitations of the \Sc~chain independently of the nanotube. Remarkably, in the use of spin peapods for quantum technologies, the main function of the nanotubes will be to give mechanical support for the endohedral fullerenes and to protect them from the enviroment, rather than to provide controlled interactions between the spins. An endohedral fullerene peapod thus provides a candidate nanostructure for spin-chain quantum computing~\cite{Ben04, Fit07}.

This work is part of QIP IRC. We thank the EPSRC's Materials Chemistry Consortium (portfolio grant EP/D504872) and MML, Oxford for providing the computing facilities. L.G. is supported by the Clarendon Fund and St. Anne's College, Oxford. J.H.J. acknowledges support from the UK MOD and Wolfson College, Oxford. G.A.D.B. thanks EPSRC for a Professorial Research Fellowship. We thank S. Benjamin for discussions.

\end{document}